\documentstyle[12pt]{article}

\makeatletter
\@addtoreset{equation}{section}
  
\makeatother

\author{Jos\'e L. Cort\'es\thanks{e-mail: cortes@dedalo.unizar.es}
 and Mikhail S. Plyushchay\thanks{On leave of absence from the
Institute for High Energy Physics, Protvino, Russia;
e-mail: mikhail@dedalo.unizar.es}\\
[1ex]{\it Departamento de F\'{\i}sica Te\'orica, Facultad de Ciencias}\\
[0.5ex]{\it Universidad de Zaragoza, 50009 Zaragoza, Spain}}

\title{ANYONS: MINIMAL AND EXTENDED FORMULATIONS}

\date{To appear in {\it Mod. Phys. Lett.} {\bf A}}

\begin{document}

\maketitle

\begin{abstract}
The minimal (reduced) and extended canonical formulations for
(2+1)-dimensional fractional spin particles are considered.
We investigate the relationship between them, clearing up the meaning
of the coordinates for such particles, and analyse the related
question of correlation between spin and momentum.
The classical lagrangian corresponding
to the extended canonical formulation is constructed,
and its gauge symmetries are identified.
\end{abstract}

\newpage
\section{Introduction}
Besides the generally accepted approach to the description of anyons \cite{any}
involving the statistical Chern-Simons U(1) gauge field \cite{sem}-\cite{rev2},
there is, in principle, another possibility to describe such
(2+1)-dimensional fractional spin particles.
It consists in using multi-valued representations of the Lorentz
group $SO(2,1)$ \cite{p5}--\cite{p*}, or infinite-dimensional
unitary representations of its universal covering group
$\overline{SL(2,R)}$  \cite{p*}--\cite{cor}. The latter case
is based on the use of some internal spin vector variables at the
classical level which lead to the above mentioned representations
after quantization.
Moreover, one can try to formulate the theory proceeding
from the classical description of relativistic particles with arbitrary spin
in terms of noncommuting (in the sense of brackets)
coordinates, not using additional spin variables at all
\cite{monop,jac,jn2}.

The present paper is devoted to the investigation of the
relationship between the latter minimal classical formulation and
the extended formulation corresponding to the
approach that uses infinite-dimensional
unitary representations of $\overline{SL(2,R)}$.
In particular, we shall investigate the correlation between
spin and momentum for a particle in 2+1 dimensions.
This question is important for anyon physics:
recently the property of parallelness of spin and momentum has been used
in Ref. \cite{elm}
as a guiding principle for the introduction of the electromagnetic
interaction of fractional spin particles within the framework
of the minimal formulation\footnote{It has been shown \cite{elm}
that the value of the gyromagnetic ratio for charged anyons,
previously obtained in a standard approach involving
the statistical Chern-Simons gauge field \cite{gfac}, is essentially
a reflection of the parallelness of spin and momentum.}.

The paper is organized as follows.
In sect. 2 proceeding from the general reasonings,
we introduce the minimal (reduced) and extended canonical
formulations for a (2+1)-dimensional particle with arbitrary spin
and then investigate in detail the first case.
Here we consider the question of correlation between spin and momentum
and discuss the quantization of the minimal formulation.
Sect. 3 is devoted to the extended formulation.
First we trace its relationship to the minimal formulation
and then discuss the lagrangian approach corresponding to it,
identifying the local gauge symmetries of the classical action.
Section 4 is devoted to concluding remarks.

\section{Minimal formulation}

As it is well known \cite{rev1,rev2},
in 2+1 dimensions the spin is a pseudoscalar variable.
So, at the classical level, within the canonical approach
a particle with fixed mass $m>0$ and spin $s\neq 0$
has the same number of degrees of freedom as a scalar particle,
and we can describe it in the following way.
Introduce the coordinate $x_{\mu}$ and momentum $p_{\mu}$ variables,
considering $p_{\mu}$, $\mu=0,1,2,$ as the components of the
energy-momentum vector being the generator of space-time translations.
Therefore, we postulate the classical brackets:
\begin{eqnarray}
\{p_{\mu},p_{\nu}\}&=&0,
\label{pp}\\
\{x_{\mu},p_{\nu}\}&=&\eta_{\mu\nu},
\label{xp}
\end{eqnarray}
where $\eta_{\mu\nu}={\rm diag} (-,+,+)$.  Then the total angular
momentum for a spinning particle can be taken in the form generalizing
that for the scalar particle:
\begin{equation}
{\cal J}_{\mu}=-\epsilon_{\mu\nu\lambda}x^{\nu}p^{\lambda} +J_{\mu}.
\label{calj}
\end{equation}
Here ${\cal J}_{\mu}$ is the vector dual to the
total angular momentum tensor ${\cal J}^{\nu\lambda}$,
${\cal J}_{\mu}=-\frac{1}{2}\epsilon_{\mu\nu\lambda}{\cal J}^{\nu\lambda}$, and
$\epsilon_{\mu\nu\lambda}$ is a totally antisymmetric tensor,
$\epsilon^{012}=1$.
${\cal J}_{\mu}$ together with
$p_{\mu}$ are the generators of the classical Poincar\'e
group, and so, brackets (\ref{pp}) should be supplemented with the brackets
\begin{eqnarray}
\{{\cal J}_{\mu},{\cal J}_{\nu}\}&=&-\epsilon_{\mu\nu\lambda}{\cal
J}^{\lambda},
\label{caljj}\\
\{{\cal J}_{\mu},p_{\nu}\}&=&-\epsilon_{\mu\nu\lambda}p^{\lambda}.
\label{caljp}
\end{eqnarray}
The condition that the particle has a fixed mass can be taken into
account by means of the constraint
\begin{equation}
p^{2}+m^{2}\approx 0
\label{p2m2}
\end{equation}
fixing the value of the classical analog of the corresponding Casimir operator
of the quanum mechanical Poincar\'e group $\overline{ISO(2,1)}$.
The second term $J_{\mu}$ in (\ref{calj})
takes into account the nontrivial spin of the particle $s\neq 0$,
and one can consider two different cases.
In the first case we can suppose that $J_{\mu}$ depends
on some $2n$ internal phase space variables,
independent on $x_{\mu}$ and $p_{\mu}$,
i.e. $\{J_{\mu},x_{\nu}\}=\{J_{\mu},p_{\nu}\}=0$, and
that the coordinates of the particle $x_{\mu}$ are commuting in the sense
of the brackets,
\begin{equation}
\{x_{\mu},x_{\nu}\}=0.
\label{xx}
\end{equation}
Then $J_{\mu}$, $\mu=0,1,2$, have to form the algebra of
Lorentz generators
\begin{equation}
\{J_{\mu},J_{\nu}\}=-\epsilon_{\mu\nu\lambda}J^{\lambda},
\label{jj}
\end{equation}
and the constraint
\begin{equation}
pJ -sm\approx 0
\label{pjsm}
\end{equation}
will fix the value of particle spin in correspondence
with  relations  (\ref{calj}), (\ref{p2m2}) and
the definition
\begin{equation}
S=\frac{p^{\mu}{\cal J}_{\mu}}{\sqrt{-p^{2}}}
\label{spin}
\end{equation}
for the classical analog of the spin Casimir operator of
the Poincar\'e group.
Since the first class constraint (\ref{pjsm}) cancels
one phase space degree of freedom (2 variables), we have to supplement
it with the corresponding number of first and (or) second class constraints
to cancel the remaining  $2(n-1)$ internal phase space variables. We shall
consider in detail an example
with additional internal phase space variables in the next section and
now turn to the minimal formulation involving only phase space
variables $x_{\mu}$ and $p_{\mu}$.
In this case due to eqs. (\ref{calj}) and (\ref{spin}),
 $J_{\mu}$ must have the form
\begin{equation}
J_{\mu}=-se^{(0)}_{\mu}+J^{(i)}
e^{(i)}_{\mu},
\label{jse}
\end{equation}
where we have introduced the triad
$e^{(\alpha)}_{\mu}=e^{(\alpha)}_{\mu}(p)$, $\alpha=0,1,2$,
$e^{(0)}_{\mu}=p_{\mu}/\sqrt{-p^{2}},$
\begin{equation}
e^{(\alpha)}_{\mu}\eta_{\alpha\beta}e^{(\beta)}_{\nu}=\eta_{\mu\nu},\quad
e^{(\alpha)}_{\mu}\eta^{\mu\nu}e^{(\beta)}_{\nu}=\eta^{\alpha\beta},\quad
\epsilon^{\mu\nu\lambda}e^{(0)}_{\mu}e^{(1)}_{\nu}e^{(2)}_{\lambda}=1.
\label{ee}
\end{equation}
Due to eqs. (\ref{xp}) and (\ref{caljp}), $J^{(i)},$ $i=1,2,$ can depend
only on $p_{\mu}$ .
Therefore, the vector
${\cal J}_{\mu}$ can be presented in the equivalent form:
\begin{equation}
{\cal
J}_{\mu}=-\epsilon_{\mu\nu\lambda}\tilde{x}{}^{\nu}p^{\lambda}-se^{(0)}_{\mu},
\label{caljn}
\end{equation}
where
\begin{equation}
\tilde{x}_{\mu}=x_{\mu}-\epsilon^{0ij}\frac{J^{(i)}}{\sqrt{-p^{2}}}e^{(j)}_{\mu}
={x}_{\mu}+ {\frac{1}{p^{2}}\epsilon_{\mu\nu\lambda}p^{\nu}}J^{\lambda}.
\label{tx}
\end{equation}
The elongated variables $\tilde{x}_{\mu}$
have the same brackets with $p_{\mu}$ as $x_{\mu}$, whereas
the form of brackets
(\ref{caljj}) partially fixes the form of the brackets between
different components $\tilde{x}_{\mu}$:
\begin{equation}
\{\tilde{x}_{\mu},\tilde{x}_{\nu}\}=\epsilon_{\mu\nu\lambda}R^{\lambda},\quad
R_{\mu}=-s\frac{p_{\mu}}{(-p^{2})^{3/2}}+R^{(i)}e^{(i)}_{\mu},
\label{txx}
\end{equation}
where $R^{(i)}$ are arbitrary functions of $p_{\mu}$.
So, for a particle
with fixed mass $m$ and spin $s$, we have
the classical Poincare algebra (\ref{pp}), (\ref{caljj}),
(\ref{caljp}), but the brackets $\{x_{\mu},x_{\nu}\}$ for the initial variables
as well as those for elongated
coordinates  (\ref{txx}) are still unfixed.

Let us introduce now the additional requirement for the
variables $\tilde{x}_{\mu}$, $\mu=0,1,2,$
to be the three components of a vector, i.e.
\begin{equation}
\{{\cal J}_{\mu},\tilde{x}_{\nu}\}=-\epsilon_{\mu\nu\lambda}
\tilde{x}{}^{\lambda}.
\label{caljtx}
\end{equation}
This requirement leads to the prescription $R^{(i)}=0$, and
completely fixes brackets (\ref{txx}):
\begin{equation}
\{\tilde{x}_{\mu},\tilde{x}_{\nu}\}=-s\epsilon_{\mu\nu\lambda}
\frac{p^{\lambda}}{(-p^{2})^{3/2}}.
\label{txxn}
\end{equation}
Therefore, in order to have the system completely defined,
we must interprete  the vector components
$\tilde{x}_{\mu}$, $\mu=0,1,2,$ as the coordinates
of the  particle with spin $s$, and in this case
the form of the angular momentum vector (\ref{caljn}) will be completely
fixed. Then the initial coordinates
$x_{\mu}$ could be considered as some auxiliary variables,
and after taking into account
eqs. (\ref{tx}) and (\ref{caljtx}), we find that
they do not form a Lorentz vector.
We shall discuss this noncovariance in the next section.

Obviuosly,
we could postulate the covariant brackets of the form (\ref{txxn})
for the initial variables $x_{\mu}$ from the very beginning, and work
in terms of only such covariant variables.
Then the addition $J_{\mu}$ would be fixed in the same covariant form as in
eq. (\ref{caljn}), i.e. we would have $J^{(i)}=0$.
But, as we shall see below,
the manifest covariance of the minimal formulation
will be inevitably lost under transition to the quantum theory even if we
work with the vector variables $\tilde{x}_{\mu}$.
Therefore, the construction of the minimal
formulation, which includes
$x_{\mu}$ as the original noncovariant variables,
is also useful for a general discussion of the formulation.
Moreover, as we shall see, such a construction
will help us to understand its relationship to the
formulations based on the addition of internal phase space variables.

Before the discussion of quantization, let us stress that only
after fixing the notion of the particle coordinates as the components
of a Lorentz vector,
the additional term in the total angular momentum vector (\ref{caljn})
will be parallell to the energy-momentum vector $p^{\mu}$.
This second term has nontrivial brackets with
the first one looking formally as the orbital angular momentum, and,
moreover, the components of neither first nor second
terms form themselves the algebra of Lorentz generators.

Concluding this section, let us demonstrate that the manifest covariance
of the formulation is lost under the transition to the quantum theory.
Indeed, the problem of quantization is reduced
to the problem of
constructing the quantum analogs of the variables $\tilde{x}_{\mu}$ having
nontrivial
brackets (\ref{txxn}). Such operators can be constructed in the following way.
Let us introduce the variables
$q_{\mu}$,
\begin{eqnarray}
&\tilde{x}_{\mu}=q_{\mu}+A_{\mu},&
\label{xqa}\\
&\{q_{\mu},q_{\nu}\}=0,\quad
\{q_{\mu},p_{\nu}\}=\eta_{\mu\nu},&\nonumber
\end{eqnarray}
by defining the ``gauge potential" $A_{\mu}=A_{\mu}(p)$ with the help of
the monopole-like relation \cite{monop,jac}:
\begin{equation}
\partial_{\mu}A_{\nu}-\partial_{\nu}A_{\mu}=
-s\epsilon_{\mu\nu\lambda}\frac{p^{\lambda}}
{(-p^{2})^{3/2}}.
\label{mono}
\end{equation}
This  relation
defines the ``gauge potential" up to a ``gauge transformation",
$ A_{\mu}\rightarrow A'_{\mu}=A_{\mu}+\partial_{\mu}f$,
and we can choose it in the form:
\begin{equation}
A_{\mu}=-s\epsilon_{\mu\nu\lambda}\frac{p^{\nu}\xi^{\lambda}}
{p^{2}+\sqrt{-p^{2}}(p\xi)},
\label{del}
\end{equation}
where $\xi^{\mu}$ is an arbitrary fixed timelike unit vector,
$\xi^{2}=-1$ (that can be taken, e.g., as $\xi^{\mu}=(1,0,0)$).
Whence we immediately conclude that due to the noncovariant form
of the ``gauge potential" (\ref{del}) and the vector character of
$\tilde{x}_{\mu}$,
the variables $q_{\mu}$ do not form a Lorentz vector having complicated
transformation properties under Lorentz boosts,
and, therefore, we loose here the manifest covariance of the formulation.
But these variables give the possibility to realize
the operators corresponding
to the classical variables $\tilde{x}_{\mu}$ in the obvious way
using the quantum analog of eq. (\ref{xqa}).

So, we see that the minimal formulation
loses its manifest covariance under transition to the quantum theory.

\section{Extended formulation}

Let us turn now to the extended formulation with
phase space variables $x_{\mu}$, $p_{\mu}$, $J_{\mu}$,
and brackets (\ref{pp}), (\ref{xp}),
(\ref{xx}), (\ref{jj}). The total angular momentum vector is given by
(\ref{calj}) and the phase space variables are constrained by
the mass and spin conditions (\ref{p2m2}), (\ref{pjsm}).
First of all, we note that the scalar $J^{2}$ lies in the center of algebra
(\ref{jj}), $\{J^{2},J_{\mu}\}=0$,
and, therefore, it can be fixed:
\begin{equation}
J^{2}=C,
\label{const}
\end{equation}
where $C$ is some real constant.
Therefore, due to eq. (\ref{const}), the number of independent internal
phase space variables will be equal to 2. So, we shall have the minimally
extended formulation.\footnote{ The most general case for the extended
formulation and its relation to the minimal formulation will be considered
elsewhere \cite{prep}.}
Then for
$C=-\alpha^{2}$, $\alpha>0$, eq. (\ref{const}) sets two
disconnected sheets of the hyperboloid:
\begin{equation}
J_{0}=\varepsilon \sqrt{\alpha^{2}+J_{i}^{2}}, \quad \varepsilon=\pm, \quad
i=1,2,
\label{2hyp}
\end{equation}
whereas in the case $C=\beta^{2}\geq 0$
it defines a one-sheet hyperboloid (degenerating into the cone at $\beta=0$).
Note here, that in correspondence with this property,
the quantization of the classical subsystem,
defined by relations (\ref{jj}) and (\ref{const}), results in the
infinite-dimensional unitary irreducible representations of the group
$\overline{SL(2,R)}$ either of the half-bounded discrete type series
$D^{\pm}_{\alpha}$ (for $C=-\alpha^{2}<0$) or of the
continuous  series (for  $C=\beta^{2}\geq 0$),
where the operator $\hat{J}_{0}$ takes correspondingly the eigenvalues
$j_{0}=\pm (\alpha+n),$ $n=0,1,2,\ldots,$ or $j_{0}=\theta+n$, $\theta \in
[0,1)$,
$n=0,\pm 1,\pm 2,\ldots$ (see Ref. \cite{sl2} for the details).

Now, let us  reveal
the relationship of the extended formulation to the minimal one.
To this end we use the complete triad (\ref{ee}) to represent the vector
$J_{\mu}$ in the form
$J_{\mu}=e^{(\alpha)}_{\mu}\eta_{\alpha\beta}J^{(\beta)}$.
The quantities
$
J^{(\alpha)}=e^{(\alpha)}_{\mu}J^{\mu}
$
satisfy the same algebra
as the initial variables $J^{\mu}$:
$
\{J^{(\alpha)},J^{(\beta)}\}=-\epsilon^{\alpha\beta\gamma}J_{(\gamma)},
$
and the spin condition (\ref{pjsm}) can be presented in the equivalent form:
\begin{equation}
J^{(0)}-s\approx 0.
\label{spi}
\end{equation}
Taking into account condition (\ref{const}), we find that the constraint
(\ref{spi}) singles out the circle $S^{1}$,
$
J^{(0)}=s,
$
$
J^{(i)}J^{(i)}=s^{2}+C,
$
as a physical subspace, i.e.
classical conditions (\ref{const}) and (\ref{spi}) are simultaneuosly
consistent
for all the values of spin $s$ when $C\geq 0$, but they are consistent only
for $s^{2}\geq \alpha^{2}$ for the case $C=-\alpha^{2}<0$.
The circle shrinks into just a one point $J^{(0)}=s$,
$J^{(i)}=0$ only when $-J^{2}=\alpha^{2}=s^{2}$. Only in this special case
the spin vector $J_{\mu}$ is parallel (in a weak sense) to the momentum vector
$p_{\mu}$.

The vector $p^{\mu}$ and
the elongated coordinates $\tilde{x}_{\mu}$ constructed from
$x_{\mu}$ according to the prescription of the same form as in eq. (\ref{tx}),
have zero brackets with
constraint (\ref{spi}) and, therefore,
they (together with $J^{(0)}$)
are the only gauge-invariant variables with respect
to the gauge transformations generated by this constraint. The
transformations are reduced simply to the
rotation for the variables $J^{(i)}$, $i=1,2$, and the gauge orbit
here is just the same circle $S^{1}$ shrinking into a point
in the above special case.

The three $J_{(\alpha)}$
can be parametrized in the following general form:
\begin{equation}
J_{(\alpha)}=(J_{(0)},J_{(i)})=\left(J_{(0)},
\sqrt{J_{(0)}^{2}+C}
\cdot n_{i}\right),\quad
n_{i}=(\cos\phi, \sin\phi),
\label{neg}
\end{equation}
where $0\leq\phi<2\pi$ and
$-\infty<J_{(0)}<\infty$ in the case $C\geq 0$,  whereas
$J_{(0)}$ can take values in the region $[\alpha,+\infty)$
or $(-\infty,-\alpha]$ when $J^{2}=-\alpha^{2}<0$.
Then, proceeding from brackets for the variables $J_{(\alpha)}$,
we find that the brackets for the independent
variables $J_{(0)}$ and $\phi$ have the form
\begin{equation}
\{\phi,J_{(0)}\}=1.
\label{phij}
\end{equation}
Now we can reduce the system to the surface defined by the spin constraint
(\ref{spi}). The reduction consists here in choosing some point on the gauge
orbit $S^{1}$. This can be done, for example,  with the help of the (local)
gauge condition:
\begin{equation}
\phi-\phi_{0}\approx 0,
\label{gauge}
\end{equation}
where $\phi_{0}$ is some fixed point, $\phi_{0}\in S^{1}$.
Note, that the reduction
can also be applied to the special case $-J^{2}=\alpha^{2}=s^{2}$
if it is considered as a limit, e.g.,
$J^{2}=-\alpha^{2}$, $s^{2}=\alpha^{2}+\epsilon^{2}$, $\epsilon\rightarrow 0$.
After calculating the Dirac brackets with the help of second class constraints
(\ref{spi}) and (\ref{gauge}), the ``spin
vector" $J^{\mu}$ is completely fixed:
\begin{equation}
J_{\mu}=-se^{(0)}_{\mu}+\gamma_{i}e^{(i)}_{\mu},
\label{red}
\end{equation}
where the constants $\gamma_{i}$ are given in terms of the angle $\phi_{0}$
via the parametrization (\ref{neg}).
The derivation of Dirac brackets is simplified if one uses
the elongated gauge invariant variables (\ref{tx}) whose
Dirac brackets $\{\tilde{x}_{\mu},\tilde{x}_{\nu}\}^{*}$
coincide with the initial brackets.
The brackets $\{\tilde{x}_{\mu},\tilde{x}_{\nu}\}$
have here the same form (\ref{txxn}) as for the variables $\tilde{x}_{\mu}$
in the minimal formulation.
The result for the Dirac brackets of the
coordinates $x_{\mu}$ is:
\begin{equation}
\{{\cal J}_{\mu},x_{\nu}\}^{*}=-\epsilon_{\mu\nu\lambda}x^{\lambda}
+\gamma_{i}e^{(i)}_{\nu}\left(-\frac{e^{(0)}_{\mu}}{\sqrt{-p^{2}}}+
e^{(j)}_{\mu}\partial^{\sigma}e^{(j)}_{\sigma}\right),
\label{jxd}
\end{equation}
\begin{equation}
\{x_{\mu},x_{\nu}\}^{*}=\epsilon_{\mu\nu\lambda}\frac{e^{(0)\lambda}}
{\sqrt{-p^{2}}}\partial^{\sigma}\left(
-\frac{1}{2}se^{(0)}_{\sigma}+\gamma_{i}e^{(i)}_{\sigma}\right).
\label{xxd}
\end{equation}
Therefore, as follows from eqs. (\ref{jxd}) and (\ref{red}),
in the general case neither coordinates $x^{\mu}$ nor ``spin" $J^{\mu}$
are any more Lorentz vectors after reduction and, besides, $J^{\mu}$ is
not parallel to the energy-momentum vector
$p^{\mu}$. Moreover, brackets (\ref{xxd}) generally have a noncovariant form.
All these noncovariant properties appear because the variables $x_{\mu}$
and $J_{\mu}$
are gauge nonivariant and they feel the noncovariance of the gauge
condition (\ref{gauge}).
Only in the case $\gamma_{i}=0$, (i.e. only when
$-J^{2}=\alpha^{2}=s^{2}$) this noncovariance in
the reduced system dissapears, and then
$x_{\mu}$ coincide with the
gauge invariant coordinates $\tilde{x}_{\mu}$ and brackets
(\ref{xxd}) turn into the covariant brackets (\ref{txxn}).
Thus, from the point of view of the reduction of the system the case
$-J^{2}=\alpha^{2}=s^{2}$ is again a special one:
only in this case $x_{\mu}$ remains a Lorentz vector,
and the brackets between different components
have a covariant (but nonzero) form. It is just only
in this case that the gauge noninvariant spin vector $J^{\mu}$ becomes
parallel to the momentum vector $p^{\mu}$ after reduction.

Concluding the discusion of the canonical approach,
we note that the Dirac brackets (\ref{jxd}),
(\ref{xxd}) for the variables $x_{\mu}$ coincide with the brackets for
$x_{\mu}$ in the minimal formulation for a choice of the addition (\ref{jse})
in the form (\ref{red}) (i.e. for $J^{(i)}=\gamma^{i}=const$).
So, the minimal formulation presented in terms of the
variables $x_{\mu}$ is nothing else than
the reduced system of the extended formulation, and the elongated covariant
coordinates $\tilde{x}_{\mu}$ given by eq. (\ref{tx}) are the gauge invariant
coordinates of the extended formulation.

Now, let us consider the lagrangian approach corresponding to
the described extended canonical system.

The brackets (\ref{jj}) for the internal variables $J_{\mu}$ can be
derived from a kinetic lagrangian
\begin{equation}
L_{kin}=-\frac{J\xi}{J^{2}+(J\xi)^{2}}\epsilon_{\mu\nu\lambda}
\xi^{\mu}J^{\nu}\dot{J}{}^{\lambda}
\label{lkin}
\end{equation}
with arbitrary fixed unit timelike vector $\xi^{\mu}$, $\xi^{2}=-1$.
The simplest way to be convinced that it is  so consists in checking the
fact
that under a Lorentz transformation of $J_{\mu}$, the kinetic term
(\ref{lkin}) is changed by a total derivative, and, therefore, it
corresponds to a Lorentz invariant term in the action.
Then, choosing $\xi^{\mu}=(1,0,0)$, and
parametrizing the variables $J_{\mu}$
as in eq. (\ref{neg}), $J_{\mu}=J_{\mu}(J_{0},\varphi)$,
one gets
$
 L_{kin}=J_{0}\cdot \dot{\varphi}.
$
{}From here we find that the brackets for independent variables
$J_{0}$ and $\varphi$ have the form (\ref{phij}) (with the substitution
$J_{(0)}$ for $J_{0}$ and $\phi$ for $\varphi$), and, therefore,
the Lagrangian (\ref{lkin}) indeed leads to the brackets (\ref{jj}).

We shall obtain the total lagrangian of the extended formulation
by adding to the kinetic term a lagrangian depending on
$\dot{x}_{\mu}$ and $J_{\mu}$ (and on auxiliary Lagrange multipliers),
which will lead to the constraints
(\ref{p2m2}) and (\ref{pjsm}), and, so, describe a system
with fixed mass and spin. Such an addition can be constructed
proceeding from the well known
lagrangian approach for a relativistic spin-$1/2$ particle.
For example, one can take the pseudoclassical lagrangian for the
Dirac particle proposed in Ref. \cite{cpv} with corresponding
substitution of odd variables by even ones and one gets
\begin{equation}
L=\frac{1}{2e}(\dot{x}_{\mu}-vJ_{\mu})^{2}-\frac{1}{2}em^{2}+
s m v+L_{kin}.
\label{ltot}
\end{equation}
Lagrangian (\ref{ltot}), with $e$ and $v$ being the  Lagrange multipliers,
leads to the mass and spin conditions (\ref{p2m2}) and (\ref{pjsm})
as secondary constraints, and, therefore, it is the lagrangian
corresponding to the extended canonical system discussed above.

The action $A=\int Ld\tau$
is invariant with respect to the reparametrizations:
$
\delta x_{\mu}=\gamma \dot{x}_{\mu},
$
$
\delta J_{\mu}=\gamma \dot{J}_{\mu},
$
$
\delta e=(\gamma e)\dot{},
$
$
\delta v=(\gamma v)\dot{},
$
$\delta L=\frac{d}{d\tau}(\gamma L)$, $\gamma=\gamma(\tau)$,
whose generator is the mass shell constraint,
and, moreover, it is invariant with respect to the transformations
generated by the spin constraint (\ref{pjsm}):
\[
\delta e=0,\quad \delta v=\dot{\rho},\quad
\delta x_{\mu}=\rho J_{\mu},\quad
\delta J_{\mu}=-\rho e^{-1}\epsilon_{\mu\nu\lambda}\dot{x}{}^{\nu}J^{\lambda},
\]
\[
\delta L=\frac{d}{d\tau}\left(\rho\left(sm+
e^{-1}\dot{x}J -J^{2}e^{-1}\left(\dot{x}J+(\dot{x}\xi)(\xi J)\right)
\cdot(J^{2}+(J\xi)^{2})^{-1}\right)\right),
\]
where $\rho=\rho(\tau)$.
The conservation of these lagrangian symmetries can be used as a
general guiding  principle for the extension of
the system to the case of its interaction with
gauge fields, e.g. with a U(1) gauge field in the
simplest case \cite{prep}.

Let us consider now the Lagrange equations of motion for
$e$ and $v$ following from (\ref{ltot}):
\begin{equation}
(\dot{x}_{\mu}-vJ_{\mu})^{2}+e^{2}m^{2}=0,\quad
\dot{x}J-vJ^{2}-s me=0.
\label{lcon}
\end{equation}
 From the second equation we get the equality
$
e=s^{-1}m^{-1}(\dot{x}J-vJ^{2}).
$
Putting it into the first equation,
we arrive at the relation
$
\dot{x}^{2}+s^{-2}(\dot{x}J)^{2}=-v(1+s^{-2}J^{2})
\cdot (vJ^{2}-2\dot{x}J).
$
{}From here we conclude that iff
\begin{equation}
-J^{2}=\alpha^{2}=s^{2},
\label{scase}
\end{equation}
there is the Lagrange constraint:
\begin{equation}
\dot{x}^{2}+s^{-2}(\dot{x}J)^{2}=\dot{x}^{2}-(\dot{x}J)^{2}\cdot
(J^{2})^{-1}=0,
\label{lagc}
\end{equation}
which means that the particle velocity vector $\dot{x}_{\mu}$ is paralell
to the spin vector $J_{\mu}$.

To conclude this section,
let us rewrite lagrangian (\ref{ltot}) in a form
revealing the speciality of the case (\ref{scase}) in a more explicite
way. To this end, we find the multiplier $v$ from the second
equation (\ref{lcon}) assuming that $J^{2}\neq 0$,
$v=(J^{2})^{-1}(\dot{x}J-sme)$,
and put it into lagrangian (\ref{ltot}).
Then we get
the following form for the total lagrangian:
\begin{equation}
L=\frac{1}{2e}\left(\dot{x}{}^{2}-(J^{2})^{-1}
(\dot{x}J)^{2}\right)+sm(J^{2})^{-1}(\dot{x}J)
-\frac{1}{2}em^{2}\left(1+s^{2}(J^{2})^{-1}\right)+L_{kin}.
\label{lspe}
\end{equation}
The term linear in $e$ dissapears from (\ref{lspe}) only
when eq. (\ref{scase}) takes place, and as a consequence,
the variation of the corresponding action through $e$
gives the Lagrange constraint (\ref{lagc}).
The spin constraint (\ref{pjsm})
appears as the primary constraint in this case,
whereas the mass-shell constraint (\ref{p2m2}) is a secondary one.

\section{Concluding remarks}
Proceeding from the classical canonical consideration of the
relativistic fractional spin particles, which does not involve
Chern-Simons U(1) gauge field constructions,
we have shown that there are (at least) two different
approaches to the description of anyons ---
with or without an explicit spin degree of freedom.
We have investigated the both cases, called here the minimal and
extended formulations, revealing their mutual relationship
and constructing the lagrangian approach corresponding to the latter
formulation.

It has been demonstrated that within the minimal approach,
the spin term in the total angular momentum vector becomes parallel
to the energy-momentum vector of the particle with nonzero arbitrary spin
only after fixing the notion of the particle
coordinates as those forming a Lorentz vector.
This fact is nontrivial. We have shown that the canonical system
for the minimal formulation
corresponds to the system appearing after
reduction of the extended canonical system to the surface defined by the
spin constraint (\ref{pjsm}). In such reduced system the initial
coordinates of the particle do not remain the components of a Lorentz vector,
but the gauge-invariant elongated coordinates
form a Lorentz vector
and their Dirac brackets coincide with the nontrivial brackets (\ref{txxn})
for covariant coordinates in the minimal formulation.

The case $-J^{2}=\alpha^{2}=s^{2}$ for the particle
with spin $s$, being contained in the
extended formulation and corresponding  to the choice of the
discrete type series of representations
$D^{\pm}_{\alpha}$ at the quantum level \cite{p*,sl2},
is a special one.
Only in this case internal spin vector $J^{\mu}$ is parallel to the
energy-momentum
vector $p^{\mu}$
(in a weak sense, or in a strong sense after reduction to the
spin constraint surface (\ref{pjsm})),
and the initial coordinates of the particle coincide
here  with the
above mentioned gauge-invariant coordinates.
Within the corresponding lagrangian formulation this case is also
a special one: only in this case there is a lagrangian
constraint in the system that prescribes the velocity of the particle
to be parallel to the time-like spin vector $J_{\mu}$.

We have demonstrated that the manifest covariance of the minimal formulation
is inevitably lost after
transition to the quantum theory. This is due to the
monopole-like equation (\ref{mono}) for the ``gauge potential"
which appears in the construction of the ``localizable"
commuting coordinates $q_{\mu}$.
Such coordinates $q_{\mu}$, having complicated transformation
properties
under Lorentz boosts, are analogs of the Newton-Wigner coordinates \cite{new}.
Though similar noncovariance takes place also for the extended formulation
at the lagrangian level
due to the presence of nondynamical arbitrary fixed
timelike unit vector $\xi_{\mu}$ in the kinetic term (\ref{lkin}),
the final hamiltonian formulation (together with the  corresponding
quantum theory \cite{p1,p*}) is, nevetherless, manifestly covariant.

$\ $

This work was partially supported by  MEC-DGICYT (Spain).


\newpage
\begin{thebibliography}{**}

\bibitem{any}
J.M. Leinaas and J. Myrheim,
{\it Nouvo Cimento} {\bf B37}, 1 (1977);\\
G.A. Goldin, R. Menikoff and D.H. Sharp,
{\it J. Math. Phys.} {\bf 22}, 1664 (1981);\\
F. Wilczek, {\it Phys. Rev. Lett.} {\bf 48}, 1144 (1982);
{\bf 49}, 957 (1982).

\bibitem{sem}
G.W. Semenoff, {\it Phys. Rev. Lett.} {\bf 61}, 517 (1988).

\bibitem{rev1}
P. Gerbert, {\it Intern. J. Mod. Phys.}  {\bf A6}, 173 (1991);\\
S. Forte, {\it Rev. Mod. Phys.} {\bf 64}, 193 (1992);\\
K. Lechner and R. Iengo, {\it Phys. Rep.} {\bf 213}, 179 (1992).

\bibitem{rev2}
F. Wilczek, {\it Fractional Statistics and Anyon Superconductivity},
(World Scientific, Singapore, 1990).

\bibitem{p5}
M.S. Plyushchay, {\it Phys. Lett.} {\bf B248}, 107 (1990).

\bibitem{fj}
S. Forte and T. Jolicoeur, {\it Nucl. Phys.} {\bf B350}, 589 (1991).

\bibitem{p*}
M.S. Plyushchay, {\it Intern. J. Mod. Phys.} {\bf A7}, 7045 (1992).


\bibitem{jac}
R. Jackiw and V.P. Nair, {\it Phys. Rev.}  {\bf D43}, 1933 (1991).

\bibitem{p1}
M.S. Plyushchay, {\it Phys. Lett.} {\bf B262}, 71 (1991).

\bibitem{p3}
D.P. Sorokin and D.V. Volkov, {\it Nucl. Phys.} {\bf B409}, 547 (1993).

\bibitem{p4}
M.S. Plyushchay, {\it Phys. Lett.} {\bf B320}, 91 (1993).

\bibitem{cor}
J.L. Cort\'es and M.S. Plyushchay,
{\it J. Math. Phys.} {\bf 35}, 6049 (1994).


\bibitem{monop}
J.F. Schonfeld, {\it Nucl. Phys.} {\bf B185}, 117 (1981).

\bibitem{jn2}
R. Jackiw and V.P. Nair,
hep-th/9403010 preprint.

\bibitem{elm}
C. Chou, V.P. Nair and A.P. Polychronakos,
{\it Phys. Lett.} {\bf B304}, 105 (1993).

\bibitem{gfac}
I.I. Kogan, {\it Phys. Lett.} {\bf B262}, 83 (1991);\\
J.L. Cort\'es, J.Gamboa and L. Vel\'azquez,
{\it Phys. Lett.} {\bf B286}, 105 (1992);
{\it Nucl. Phys.} {\bf B392}, 645 (1993).

\bibitem{prep}
J.L. Cort\'es and M.S. Plyushchay, in preparation.

\bibitem{sl2}
M.S. Plyushchay, {\it J. Math. Phys.} {\bf 34}, 3954 (1993).

\bibitem{cpv}
J.L. Cort\'es, M.S. Plyushchay and L. Vel\'azquez,
{\it Phys. Lett.} {\bf B306}, 34 (1993).


\bibitem{new}
T.D. Newton and E.P. Wigner, {\it Rev. Mod. Phys.} {\bf 21}, 400 (1949).

\end{thebibliography}
\end{document}